%% file: eprint_cipanp2015_exotics.tex
\documentclass[12pt]{article}
\usepackage{graphicx}
\usepackage{overpic}
\usepackage{amsmath}

\newcommand\pubnumber{CIPANP2015-Bian}
\newcommand\pubdate{\today}

\def\umn{School of Physics and Astronamy\\
University of Minnesota, MN 55455, USA}

\def\Title#1{\begin{center} {\Large #1 } \end{center}}
\def\Author#1{\begin{center}{ \sc #1} \end{center}}
\def\Address#1{\begin{center}{ \it #1} \end{center}}

\newcommand\pubblock{\rightline{\begin{tabular}{l} \pubnumber\\
         \pubdate  \end{tabular}}}
\newenvironment{Abstract}{\begin{quotation}  }{\end{quotation}}
\newenvironment{Presented}{\begin{quotation} \begin{center} 
             PRESENTED AT\end{center}\bigskip 
      \begin{center}\begin{large}}{\end{large}\end{center} \end{quotation}}

\input econfmacros.tex

\begin{document}
\begin{titlepage}
\pubblock

\vfill
\Title{ Studies of Charmonium-like States at BESIII}
\vfill
\Author{ Jianming Bian \\
(for the BESIII Collaboration)}
\Address{\umn}
\vfill
\begin{Abstract}

In the quark model, hadrons are dominantly bound states of quark-antiquark pairs (mesons) or three quarks (baryons), but QCD also allows hadronic states to be composed of more quarks bound together.  Recently, BESIII, Belle and LHCb have confirmed the existence of four-quark and pentaquark candidates.  These new states, along with experimentally observed resonances that do not fit well into the charmonium and bottomonium spectra, present challenges and opportunities for strong interaction theory.  In this talk, I will review results on charmonium-like exotic quark states that have been observed by the BESIII experiment at the Beijing Electron Positron Collider II (BEPCII).

\end{Abstract}
\vfill
\begin{Presented}
CIPANP 2015\\
Twelfth Conference on the Intersections of Particle and Nuclear Physics\\
Vail, Colorado, May 19--24, 2015\\
\end{Presented}
\vfill
\end{titlepage}
\def\thefootnote{\fnsymbol{footnote}}
\setcounter{footnote}{0}

\section{Introduction}

In the quark model, hadrons are dominantly bound states of mesons or baryons, which consist of two or three quarks. However, quantum chromodynamics (QCD) allows hadrons that contain no quarks or more than three quarks. Proposed exotic states include glueballs (consisting solely of gluons, with no quarks), hybrids (with quarks and excited gluon), multi-quark states (with more than three quarks), and hadron molecules (bound states of two or more hadrons like the deuteron). There is a long history of searches for these exotic hadrons, however, no solid experimental evidence was found until recent breakthroughs in the charmonium and bottomonium regions. This paper presents recent studies of charmonium-like exotic quark states at BESIII.

Charmonium is a bound state of a charmed quark and antiquark. It is one of the simplest bound states of QCD, like positronium in QED. Below the $D\bar{D}$ open-charm threshold, all expected charmonium states have been observed and $c\bar{c}$ potential models describe the spectrum very well. However, there are many missing states above open-charm threshold. In this energy region, a number of new states have been observed. These states are not obvious charmonium states, but all of them have charmonium among their decay products.They are called charmonium-like states, and they are classified in three categories, $X$, $Y$ and $Z$. $X$ states are neutral and produced in $B$ decays, $Y$ transitions and hadron machines.  $Y$ states are neutral vectors ($1^{-}$), which can be directly produced in $e^+e^-$ colliders. $Z^{\pm}$ states are charged quarkonium-like particles.

The Beijing Electron-Positron Collider II (BEPCII) at the Institute of High Energy Physics (IHEP) in Beijing is a two-ring electron-positron collider that runs in the tau-charm energy region ($E_{cm} = 2.0-4.6$ GeV) with a design luminosity of $1 \times 10^{33}$ cm$^{-2}$s$^{-1}$ at a beam energy of 1.89 GeV. The BEPCII accelerator complex consists of a 202-m-long electron-positron linac injector that accelerates the electrons and positrons to 1.3 GeV, and two storage rings with a circumference of 237.5 m, one for electrons and one for positrons. Electrons and positrons collide at the interaction point (IP), where the BESIII detector is located with a horizontal crossing angle of 11 mrad and bunch spacing of 8 ns. Each ring has 93 bunches with a design beam current of 910 mA. The BESIII detector consists of a beryllium beam pipe, a helium-based small-celled drift chamber, Time-Of-Flight counters (TOF) for particle identification, a CsI(Tl) crystal calorimeter, a superconducting solenoidal magnet with a field of 1 Tesla, and a muon identifier using Resistive Plate Counters (RPC) interleaved with layers of the magnet flux-return iron. Physics programs at BESIII include light-hadron and charmonium spectroscopy, electroweak and strong physics at the charm scale, tau-physics, $R$ value measurements, and searches for rare processes.  For the $XYZ$ studies, BESIII has accumulated about 5 fb$^{-1}$ of data, including high-statistics data samples around $\psi$(4040), $Y$(4260) and $Y$(4360). Data samples with small statistics at other energy points in the $XYZ$ energy region have been collected for line-shape studies.

\section{\boldmath X states} 
\subsection{$e^+e^-\to \gamma X(3872)\to \gamma \pi^+\pi^- J/\psi$}

\begin{figure}[h]
\begin{center}
\includegraphics[height=5cm]{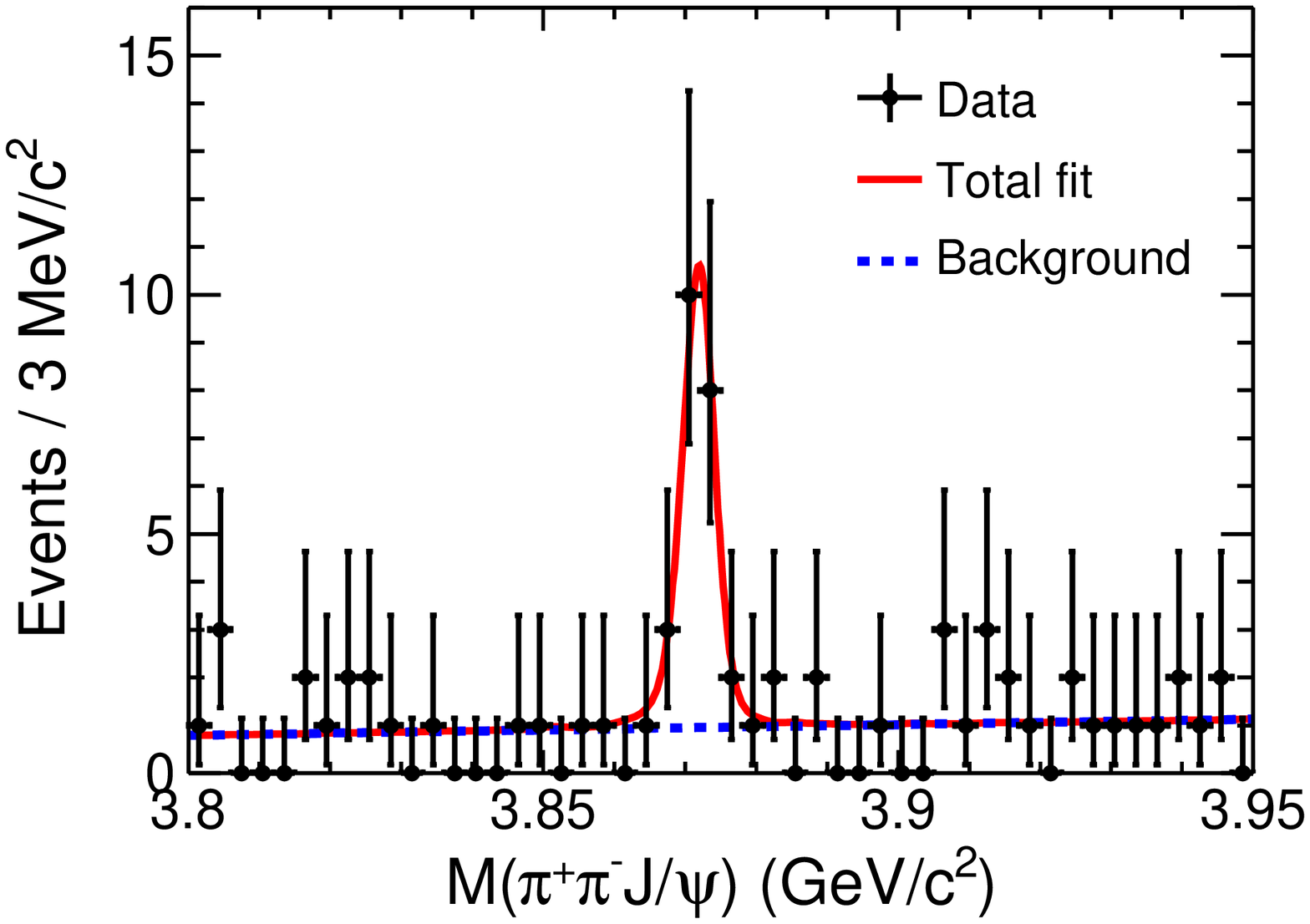}
\includegraphics[height=5cm]{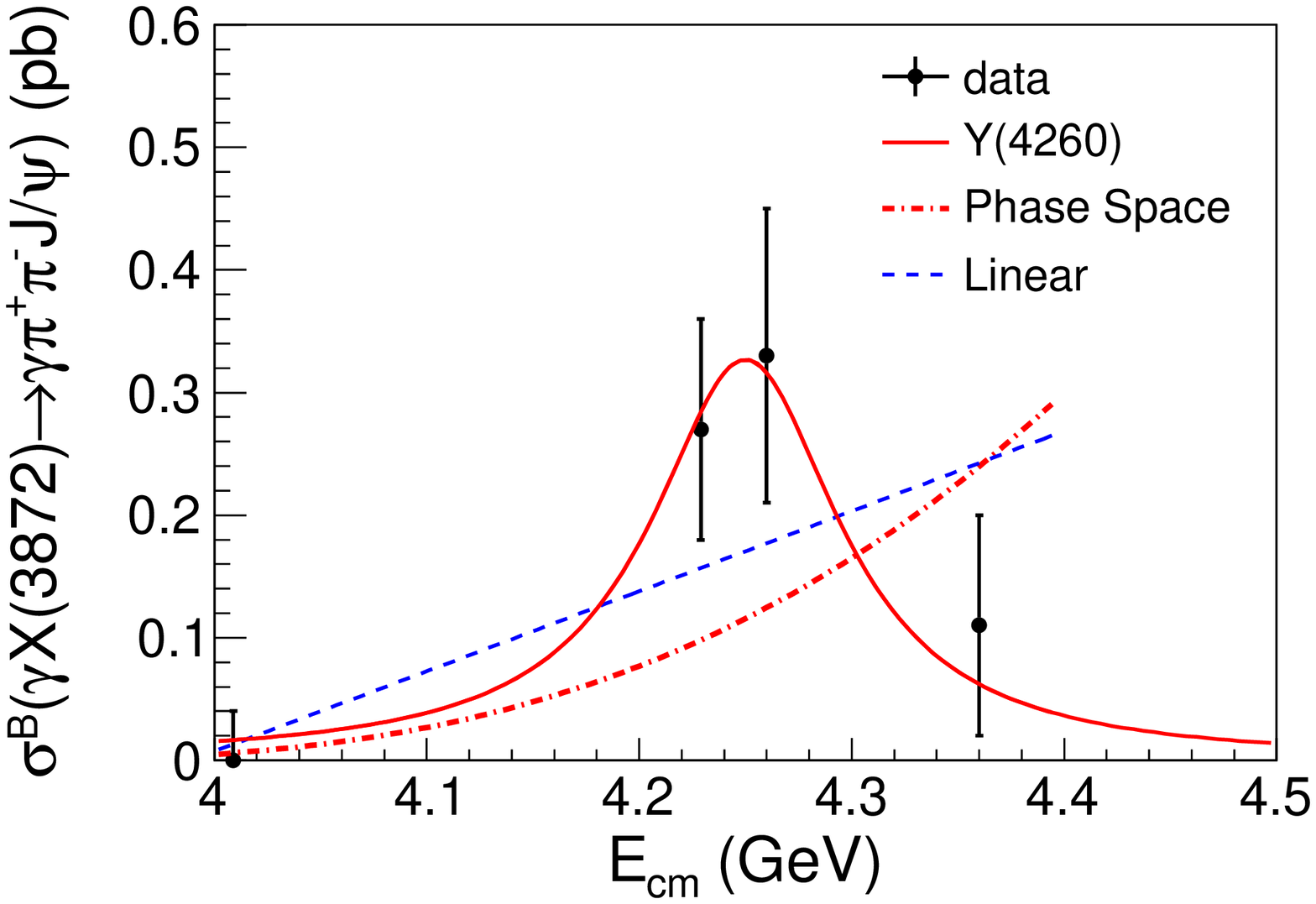}
\caption{(left) $M(\pi^+\pi^-J/\psi)$ in the process $e^+e^-\to \gamma \pi^+\pi^- J/\psi, J/\psi\to e^+e^-,\mu^+\mu^-$; (right) fit to $\sigma^B[e^+e^-\to \gamma X(3872)]\times
\mathcal{B}[X(3872)\to \pi^+\pi^- J/\psi]$} \label{fig:gx3872}
\end{center}
\end{figure}

BESIII has observed $X(3872)$ through the radiative transition $e^+e^-\to \gamma X(3872)\to \gamma \pi^+\pi^- J/\psi$, with $J/\psi$ reconstructed via the dilepton modes~\cite{Ablikim:2013dyn}. The $M(\pi^+\pi^-J/\psi)$ distribution, summing over all energy points is shown in Figure~\ref{fig:gx3872}~(left), which is used to extract the mass and yield of $X(3872)$. The statistical significance of the $X(3872)$ signal is $6.3\sigma$.  The $X(3872)$ mass is measured to be $(3871.9\pm 0.7\pm 0.2)$~MeV/$c^2$ when fitting with the width fixed to the PDG value~\cite{pdg}. From a fit with a floating width, we obtain a width of $(0.0^{+1.7}_{-0.0})$~MeV, or less than 2.4~MeV at the 90\% C.L. 

The energy-dependent Born cross-sections are measured at c.m. energies of $\sqrt{s}=4.009, 4.229, 4.260$, and 4.360 GeV, as shown in Figure~\ref{fig:gx3872}~(right). The measured cross-sections at 4.26 GeV and 4.36 GeV are an order of magnitude higher than the NRQCD calculation of continuum production. We fit these cross-sections with (1) the $Y(4260)$ resonance (parameters fixed to PDG values); (2)  the linear continuum line shape; (3) and the $E1$-transition phase space ($\propto E^3_\gamma$) factor. Goodnesses of these fits are (1) $\chi^2/{\rm ndf}=0.49/3$ (C.L. = 92\%), (2) 5.5/2 (C.L. = 6\%), and (3) 8.7/3 (C.L. = 3\%), respectively, which indicates that the $Y(4260)$ resonance describes the line shape of  $e^+e^-\to \gamma X(3872)$ better than the other two options. This observation indicates that the $X(3872)$ is produced via the radiative transition $Y(4260)\to\gamma X(3872)$. The ratio $\mathcal{B}(Y(4260)\to\gamma X(3872))/\mathcal{B}(Y(4260)\to\pi^+\pi^- J/\psi)$ is about $11\%$, which is a relatively large transition ratio. Together with the observation of $Y(4260)\to\pi^{\mp}Z_c(3900)^{\pm}$,  the BESIII observation demonstrates  some commonality in the nature of the exotics states $X(3872)$, $Y(4260)$ and $Z_c(3900)$.

\subsection{$e^+e^-\to \pi^+\pi^- X(3872) \to \pi^+\pi^- \gamma \chi_{c1}$}

\begin{figure}[h]
\begin{center}
\includegraphics[height=5cm]{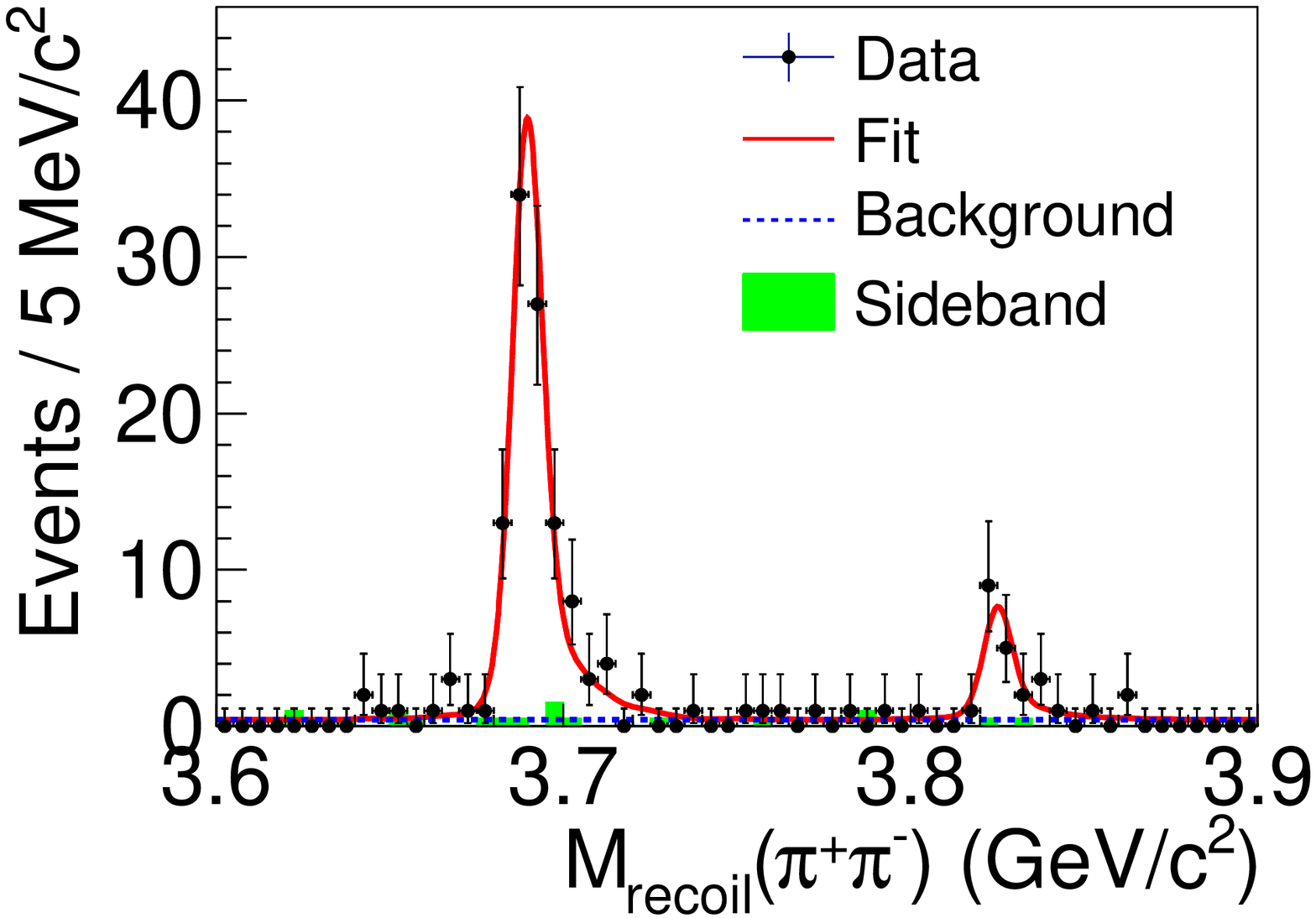}
\includegraphics[height=5cm]{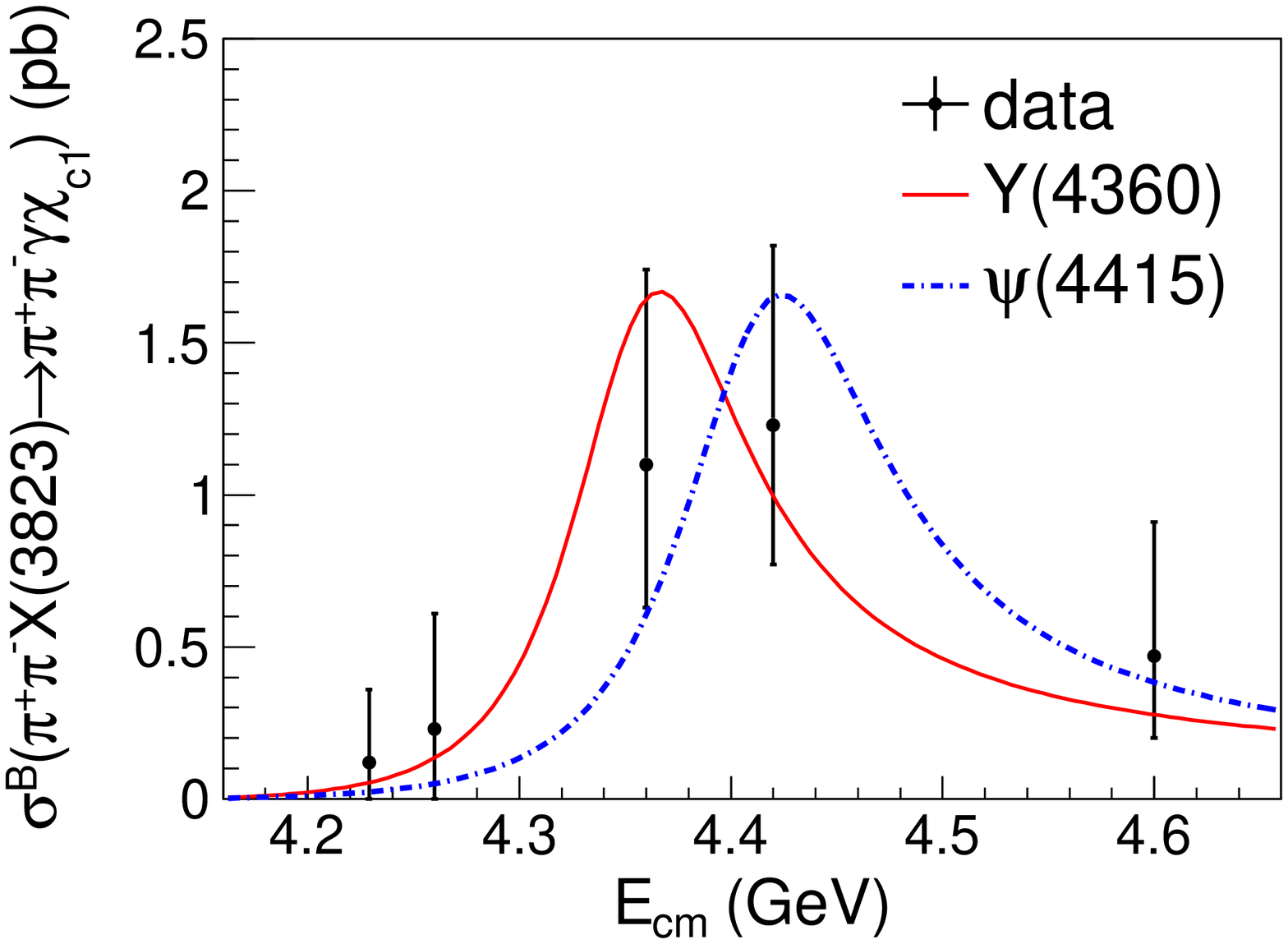}
\caption{(left) $M_{\rm recoil}(\pi^+\pi^-)$ distribution of $\gamma\chi_{c1}$ events; (right) energy-dependent  cross-sections of $\sigma^B[e^+e^-\to\pi^+\pi^- X(3823)]\times \mathcal{B}(X(3823)\to\gamma\chi_{c1})$ (dots) and the $Y(4360)$ and $\psi(4415)$ line shapes (curves).} \label{fig:pipix3823}
\end{center}
\end{figure}

BESIII observed $X(3823)$ in the process $e^+e^-\to \pi^+\pi^-X(3823) \to\pi^+\pi^-\gamma\chi_{c1}$ at c.m. energies of $\sqrt{s}=$4.230, 4.260, 4.360, 4.420, and 4.600~GeV~\cite{Ablikim:2015dlj}. Figure~\ref{fig:pipix3823} (left) shows the fit to the $\pi^+\pi^-$ recoil mass spectrum against the $\gamma\chi_{c1}$ system. The fit yields $19\pm 5$ $X(3823)$ signal events in the $\gamma\chi_{c1}$ mode,  with a statistical significance of $6.2\sigma$. The mass of $X(3872)$  is measured to be $(3821.7\pm 1.3\pm 0.7)~{\rm MeV}/c^2$. The width of $X(3872)$ is determined to be $\Gamma[X(3823)]<16$~MeV at the 90\% C.L, including systematic errors. This measurement is consistent with BELLEÕs $3.7\sigma$ evidence for $X(3823)$~\cite{belle-3d2}. The production cross-section $\sigma^{B}(e^+e^-\to\pi^+\pi^-X(3823))\times \mathcal{B}(X(3823)\to \gamma\chi_{c1})$ as a function c.m. energy is shown in Figure~\ref{fig:pipix3823}. The cross-sections are fitted with the $Y(4360)$ and $\psi(4415)$ line shapes, with their resonance parameters fixed to the PDG values. Both the $Y(4360)$ and $\psi(4415)$ hypotheses are accepted at a 90\% C.L. 

$X(3823)$ is a good candidate for the $\psi(1\,^3D_2)$ charmonium state. The mass of $X(3823)$ agrees with the $\psi(1\,^3D_2)$ prediction, $3.810-3.840$ GeV by potential models. The process $\psi(1\,^3D_2)\to D\bar{D}$ violates parity and the predicted mass of $\psi(1\,^3D_2)$ is below $D\bar{D}^*$ threshold, so $\psi(1\,^3D_2)$ does not decay to open-charm final states and its width is expected to be narrow, in agreement with the observation. For the other two charmonium candidates in this mass region, $\psi(1\,^1D_2)$ and $\psi(1\,^3D_3)$, the decay mode $\gamma\chi_{c1}$ is suppressed. In addition, the ratio $\mathcal{B}[X(3823)\to \gamma\chi_{c2}]/\mathcal{B}[X(3823)\to \gamma\chi_{c1}]$ has an upper limit of 0.42 at the 90\% C.L., which also agrees with expectations for the $\psi(1\,^3D_2)$ state, $\sim 0.2$.~\cite{ratio}.

\section{\boldmath $Y$ states} 

$Y$ states are a family of vectors observed in $e^+e^-$ colliders. In initial-state radiative processes, BaBar, Belle and CLEO observed $Y(4260)\to\pi^+\pi^- J/\psi$, $Y(4360)\to \pi^+\pi^- \psi'$ and $Y(4660)\to\pi^+\pi^-\psi'$~\cite{Aubert:2005rm, Wang:2007ea, He:2006kg}. Properties of these $Y$ states are different from conventional vector charmonium states, including their small decay rates to open charm. Between 4 and 4.7 GeV, at most five $1^{--}$ states are expected in the charmonium family (3S, 2D, 4S, 3D, 5S). However, including these $Y$ states, at least seven particles have been established. Possible explanations include hybrids, molecular states, hadrocharmonium, threshold effects, and FSI effects. Although the exact nature of the $Y$ family is still unclear, we know that $Y(4260)$, $Y(4360)$ and $Y(4660)$ are narrow and similar to each other. At BESIII, the vector $\psi/Y$ states can be produced directly in the electron-positron collision, so we measure cross sections of charmonium productions around the $Y$-family region to look for connections between the $Y$-family and traditional charmonium states.

\subsection{\boldmath $e^+e^-\to \pi^+\pi^- h_c$}

\begin{figure}[htbp]
\begin{center}
\includegraphics[width=10cm]{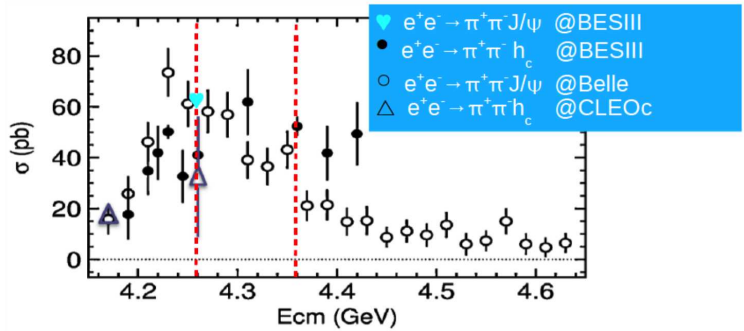}
\caption{Cross sections of $h_c$ and $J/\psi$ productions around the $Y$ region. Red dashed lines represent nominal masses of $Y(4260)$ and $Y(4360)$.} \label{fig:xsec}
\end{center}
\end{figure}

BESIII has measured $e^+e^-\to \pi^+\pi^- h_c$ at 13 c.m. energies from 3.900 to 4.420 GeV~\cite{Ablikim:2013wzq}, with $h_c$ reconstructed via $h_c\to\gamma\eta_c$, $\eta_c\to hadrons$ (16 exclusive modes). Cross-sections of $\pi^+\pi^-J/\psi$ and $\pi^+\pi^- h_c$ measured by Belle, BESIII and CLEOc are shown in Figure~\ref{fig:xsec}. Cross-sections of the two channels are of the same order of magnitude, but their line shapes are different. Cross-sections of $\pi^+\pi^-J/\psi$  as a function of c.m. energy peaks at 4.260 GeV, the nominal mass of Y(4260), while the line shape of $\pi^+\pi^- h_c$ has a local maximum around 4.230 GeV  and a broad structure at higher energy. Whether the different line shapes are caused by a new resonance or interference between $Y(4260)$ and $Y(4360)$ is still under study.

\subsection{\boldmath $e^+e^-\to \omega\chi_{c0}$}

\begin{figure}[htbp]
\begin{center}
\includegraphics[width=0.6\textwidth]{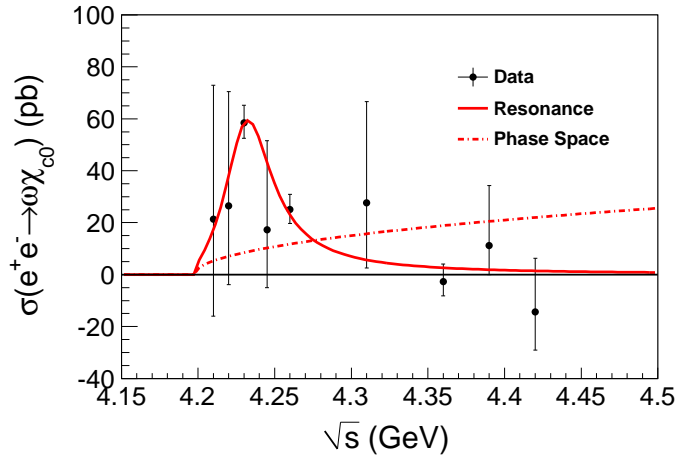}
\caption{Fit to $\sigma(e^+e^-\to\omega\chi_{c0})$ with a
resonance (solid curve), or a phase space term (dot-dashed curve).} \label{CS-BW-float}
\end{center}
\end{figure}

Based on data samples collected between $\sqrt{s}=4.21$ and 4.42~GeV, the $e^+e^-\to \omega\chi_{c0}$ process was observed at $\sqrt{s}=4.23$ and 4.26~GeV by BESIII~\cite{Ablikim:2014qwy}. Born cross-sections of $e^+e^-\to \omega\chi_{c0}$ were determined to be $(55.4\pm 6.0\pm 5.9)$ and $(23.7\pm 5.3\pm3.5)$~pb, respectively. For other c.m. energies and processes $e^+e^-\to \omega\chi_{c1,2}$, no significant signals were found. The Born cross sections of $e^+e^-\to \omega\chi_{c0}$ as a function of c.m. energy is shown in Figure~\ref{CS-BW-float}. The line shape appears to peak at around 4.230~GeV/$c^2$. Fitting with a single resonance, $\Gamma_{ee} \mathcal{B} (\omega\chi_{c0})$, the mass and the width of the resonance are determined to be $(2.7\pm 0.5\pm 0.4)$~eV, $(4230\pm 8\pm 6)$~MeV/$c^2$ and $(38\pm 12\pm 2)$~MeV, respectively. This indicates that $\omega\chi_{c0}$ signals are not purely from $Y(4260)$.

\subsection{\boldmath $e^+e^-\to \eta J/\psi$, $e^+e^-\to \eta' J/\psi$ and $e^+e^-\to \pi^0\eta J/\psi$}

\begin{figure}[h]
\begin{center}
\includegraphics[height=5cm]{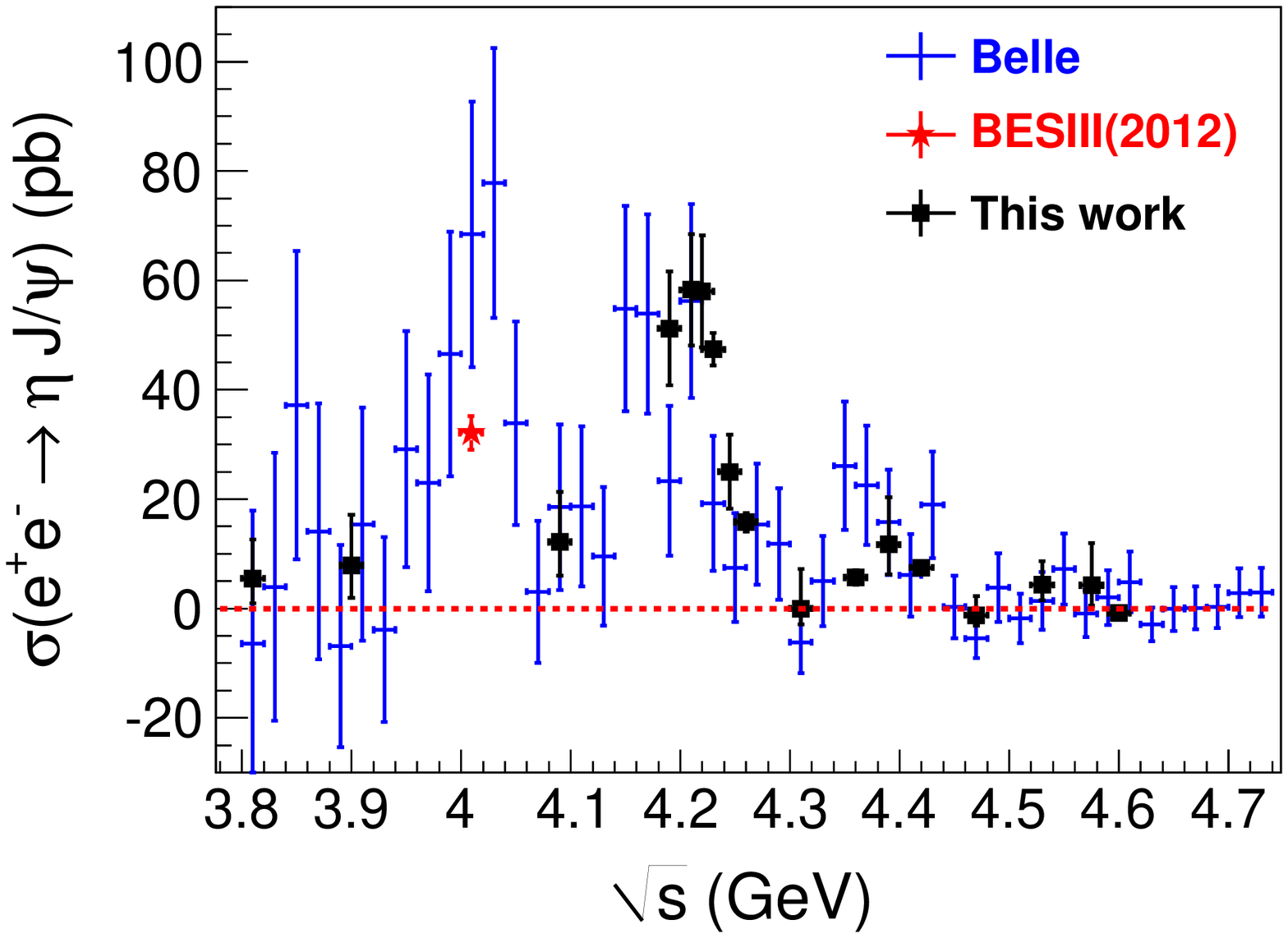}
\includegraphics[height=5cm]{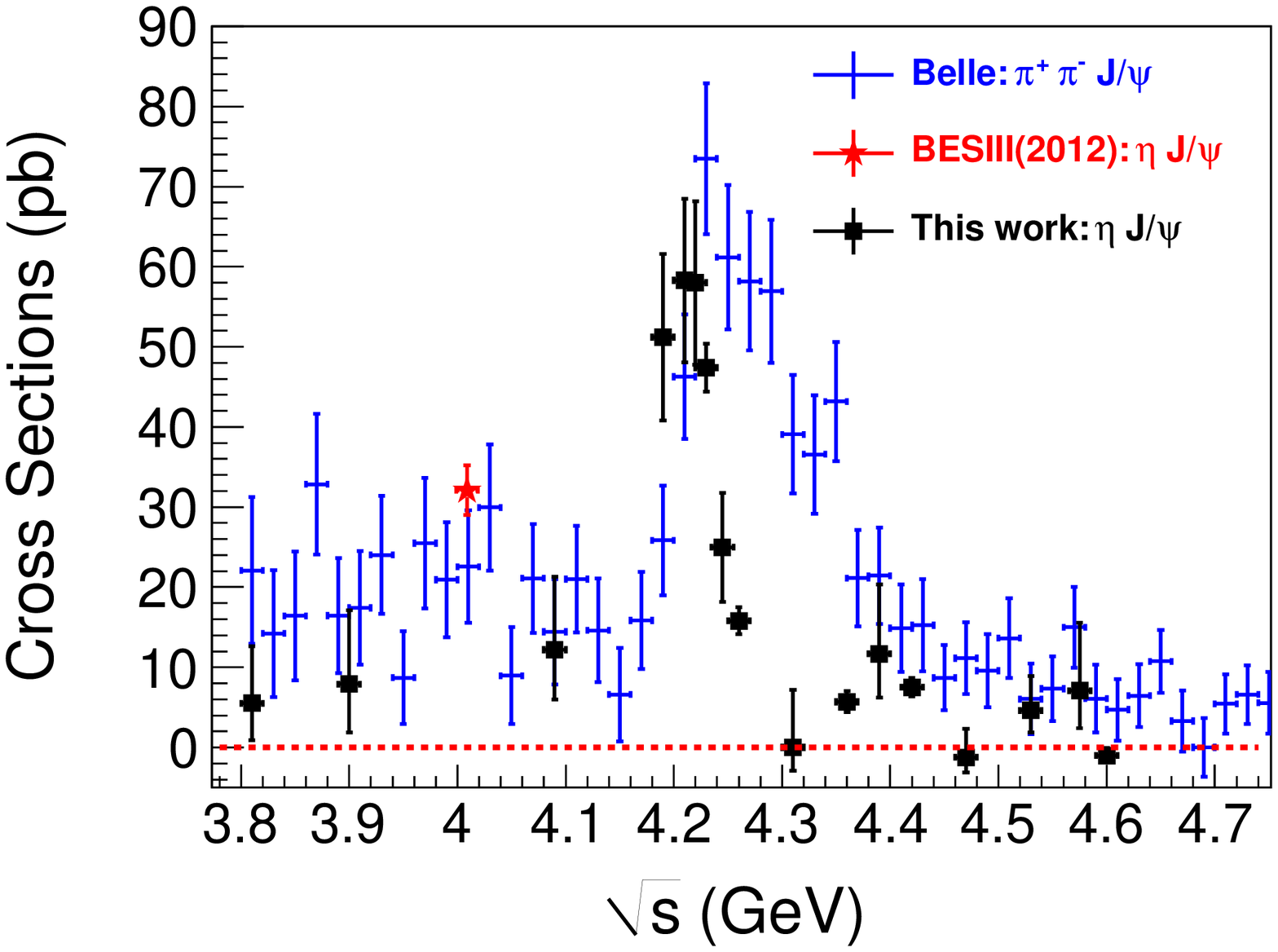}
\caption{Comparisons of Born cross sections measured by BESIII and Belle: (left) $e^{+}e^{-} \to \eta J/\psi$; (right) $e^{+}e^{-}\to \pi^{+}\pi^{-} J/\psi$.  Errors are statistical only for Belle results, and are combined statistical and systematic for BESIII results.}\label{BES_BELLE}
\end{center}
\end{figure}

Using data samples collected at energies from 3.81 to 4.60~GeV, BESIII measured the Born cross-section of $e^+e^-\to \eta J/\psi$ as a function of c.m. energy \cite{etajpsi}. As shown in Figure~\ref{BES_BELLE} (left), BESIII result is consistent with Belle's measurements~\cite{4040,belle}, but has greater statistical significance.  The Born cross-sections of $e^+e^-\to \eta J/\psi$ from BESIII are also compared to those of $e^{+}e^{-} \to \pi^{+}\pi^{-} J/\psi$ from Belle~\cite{belley_new}, as shown in Fig.~\ref{BES_BELLE} (right). Significant differences can be observed between these two processes, indicating that the $Y$ states in this energy region have different couplings to charmonium states. In addition, evidence for $e^+e^-\to \eta' J/\psi$ is also observed at 4.23 GeV and 4.36 GeV in a preliminary BESIII measurement. No evidence is found for the isospin-violating process  $Y(4260)\to \pi^0\eta J/\psi$ \cite{Ablikim:2015xfo}. Upper limits of this process are still well above the sensitivity requirement for distinguishing different models.


\subsection{\boldmath Search for $e^+e^-\to \gamma \chi_{cJ}$ and $e^+e^-\to \gamma Y(4140)$}
Using data samples collected at $\sqrt{s}$ = 4.009, 4.23, 4.26, and 4.36~GeV, BESIII searched for the processes $e^+e^-\to \gamma\chi_{cJ}, \chi_{cJ}\to \gamma J/\psi, J/\psi \to \mu^+\mu^-$ \cite{Ablikim:2014hwn}. Evidence for $e^+e^-\to \gamma\chi_{c1}$ and $e^+e^-\to \gamma\chi_{c2}$ was obtained with statistical significances of 3.0$\sigma$ and 3.4$\sigma$, respectively. No evidence of $e^+e^-\to \gamma\chi_{c1}$ was observed. 

Evidence for $Y(4140)$ was reported by CDF~\cite{Aaltonen:2009tz}. This state has a positive C-parity and can be searched for through radative transitions from other vectors. BESIII searched for $Y(4140)$ in the process $e^+e^- \to \gamma \phi J/\psi$ at $\sqrt{s} = 4.23$, 4.26, and 4.36~GeV but no significant $Y(4140)$ signal was observed~\cite{Ablikim:2014atq}. The upper limits of $\sigma[e^+e^- \to \gamma Y(4140)] \times \mathcal{B}(Y(4140) \rightarrow \phi J/\psi)$ at the 90\% C.L. were set to 0.35, 0.28, and 0.33~pb at $\sqrt{s} = 4.23$, 4.26, and 4.36~GeV, respectively. Comparing these upper limits with the result of $e^+e^-\to\gamma X(3872)$ from BESIII~\cite{Ablikim:2013dyn}, the ratio ${\sigma[e^+e^-\to\gamma Y(4140)]}/{\sigma[e^+e^- \to\gamma X(3872)]}$ is estimated to be $\sim 0.1$ or less at $\sqrt{s} = 4.23$ and 4.26~GeV.

\section{\boldmath $Z_c$ states} 

\begin{figure}[!htb]
\centering
\begin{overpic}[width=0.24\linewidth, height=2.55 cm]{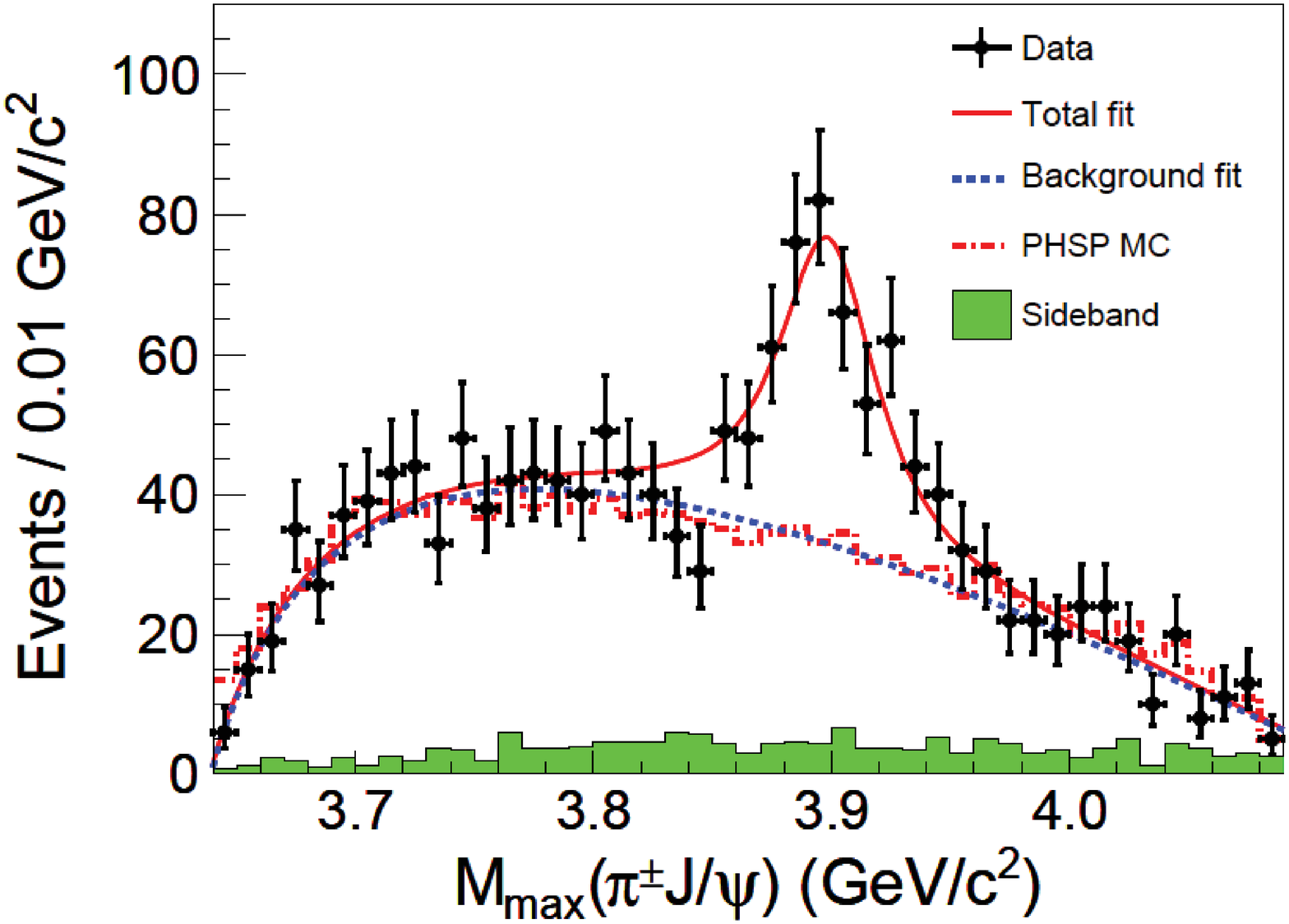}
\end{overpic}
\begin{overpic}[width=0.24\linewidth, height=4 cm]{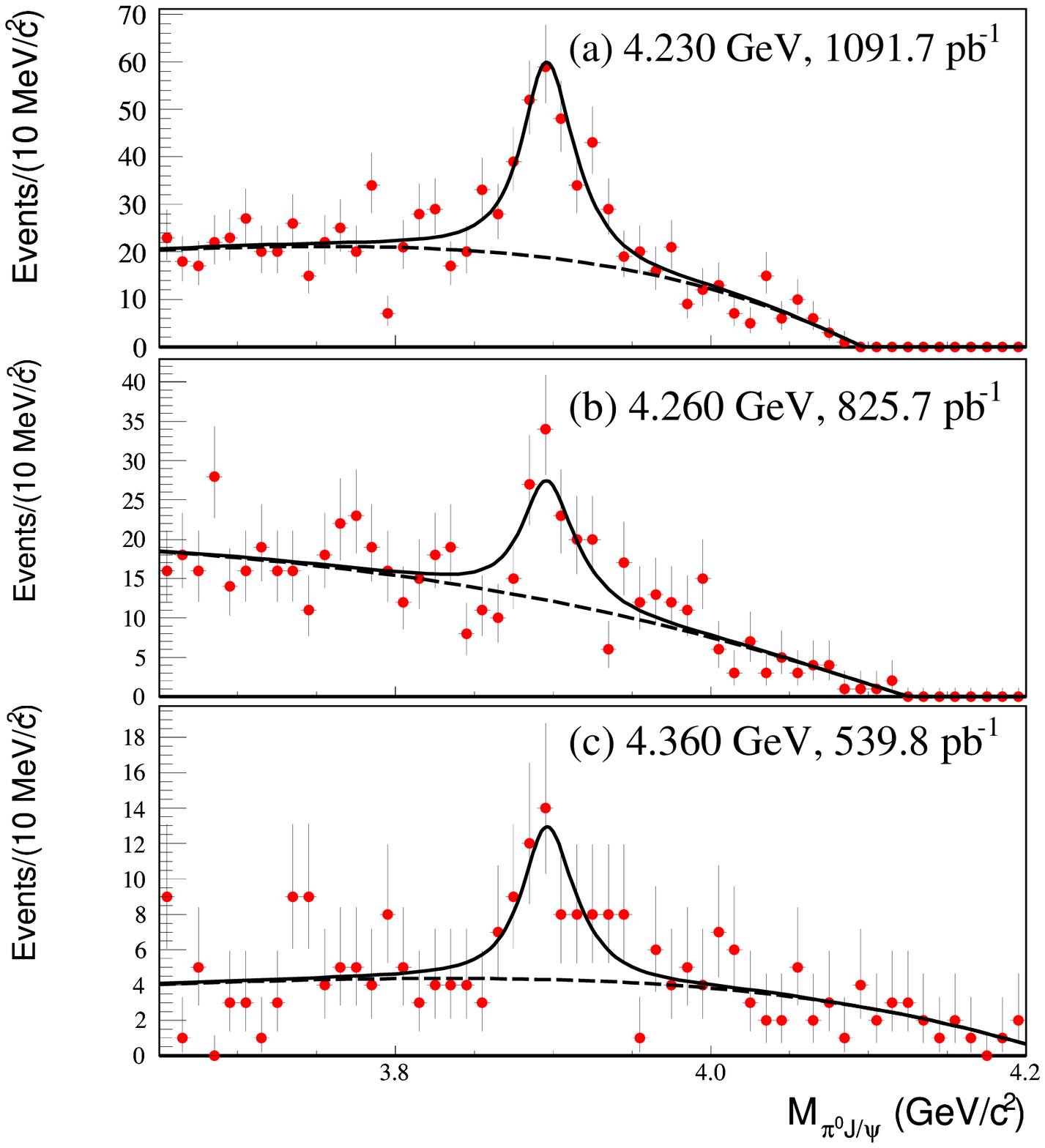}
\end{overpic}
\begin{overpic}[width=0.24\linewidth, height=2.55 cm]{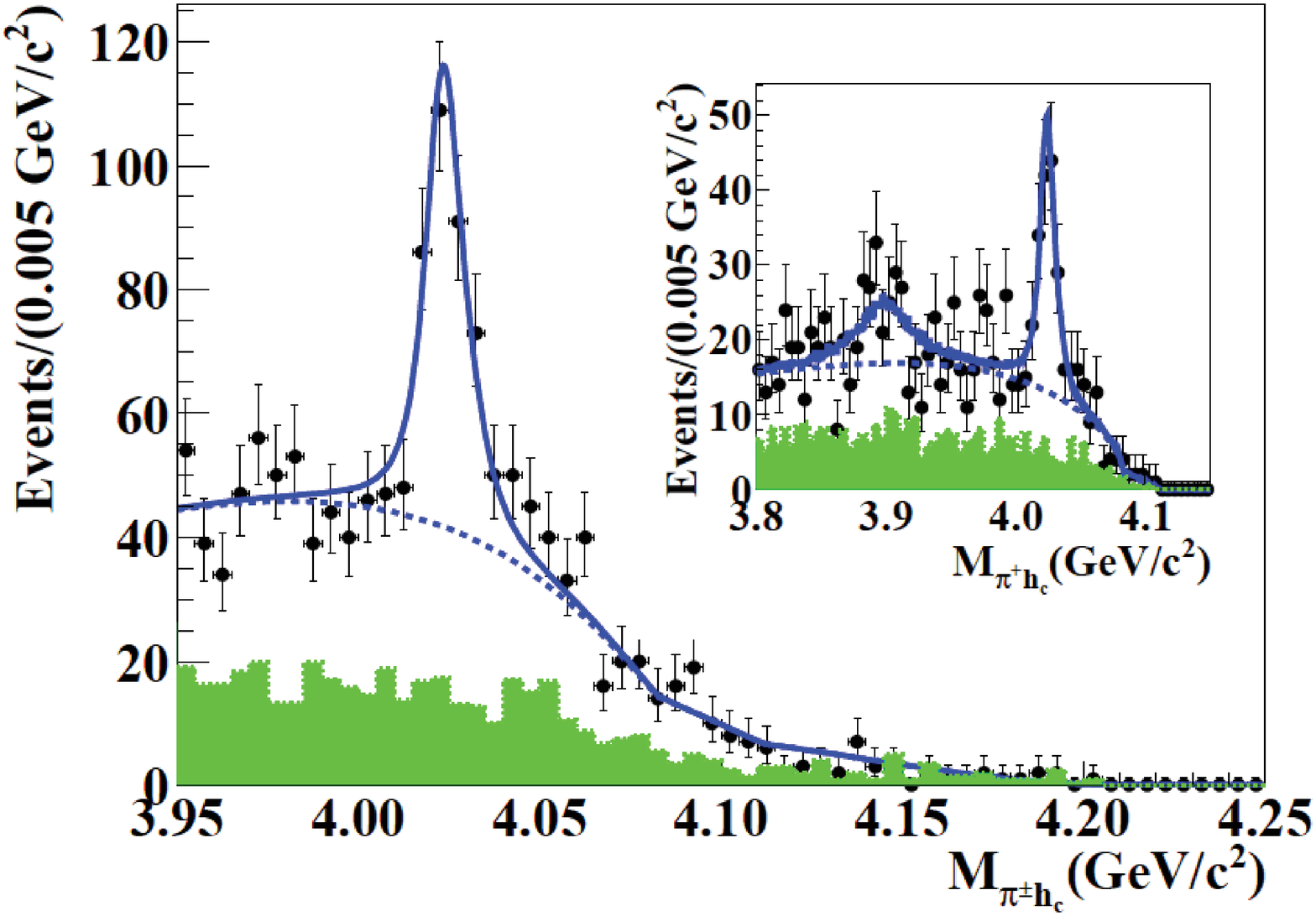}
\end{overpic}
\begin{overpic}[width=0.24\linewidth, height=2.55 cm]{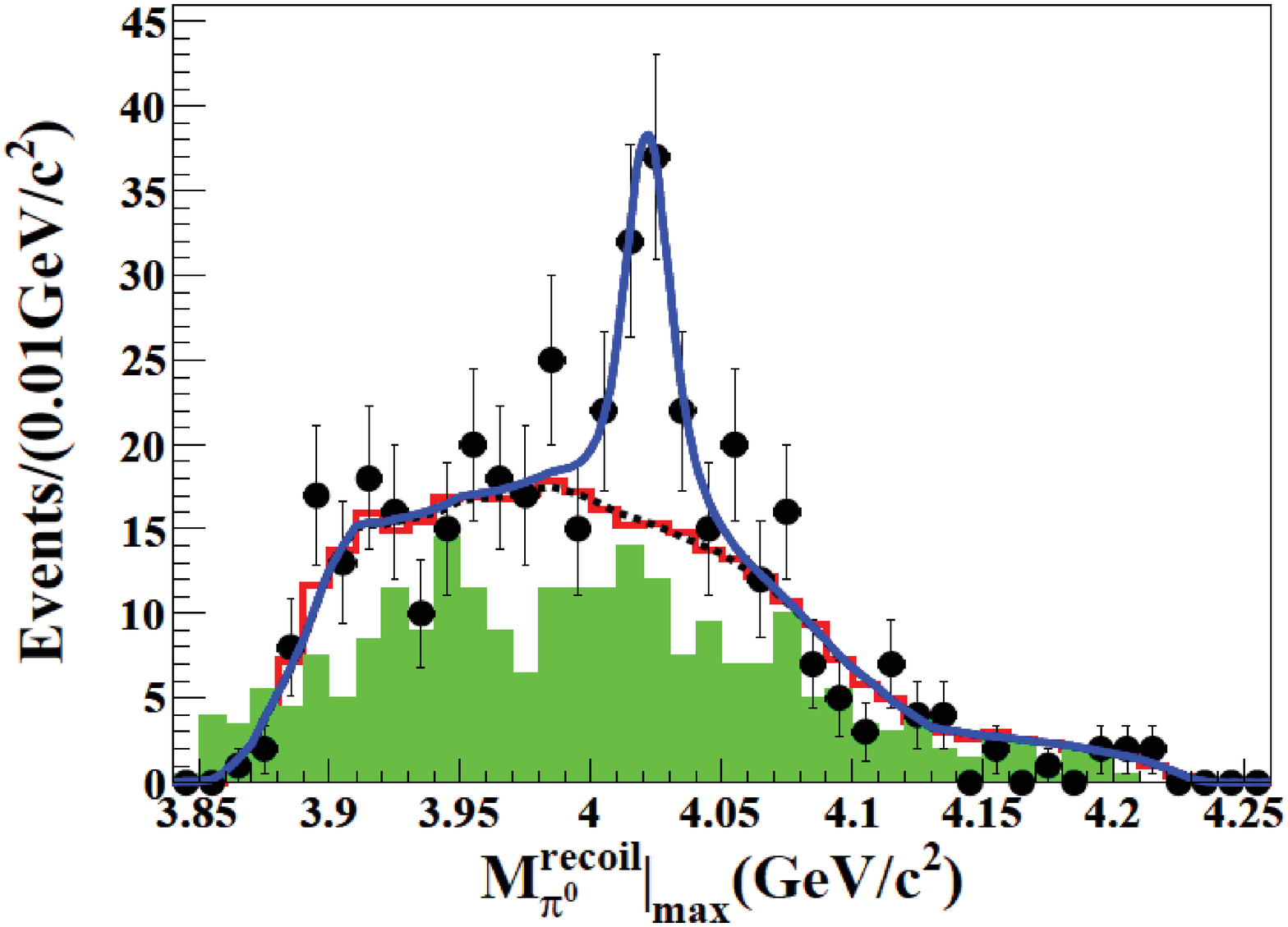}
\end{overpic}
\begin{overpic}[width=0.24\linewidth, height=2.55 cm]{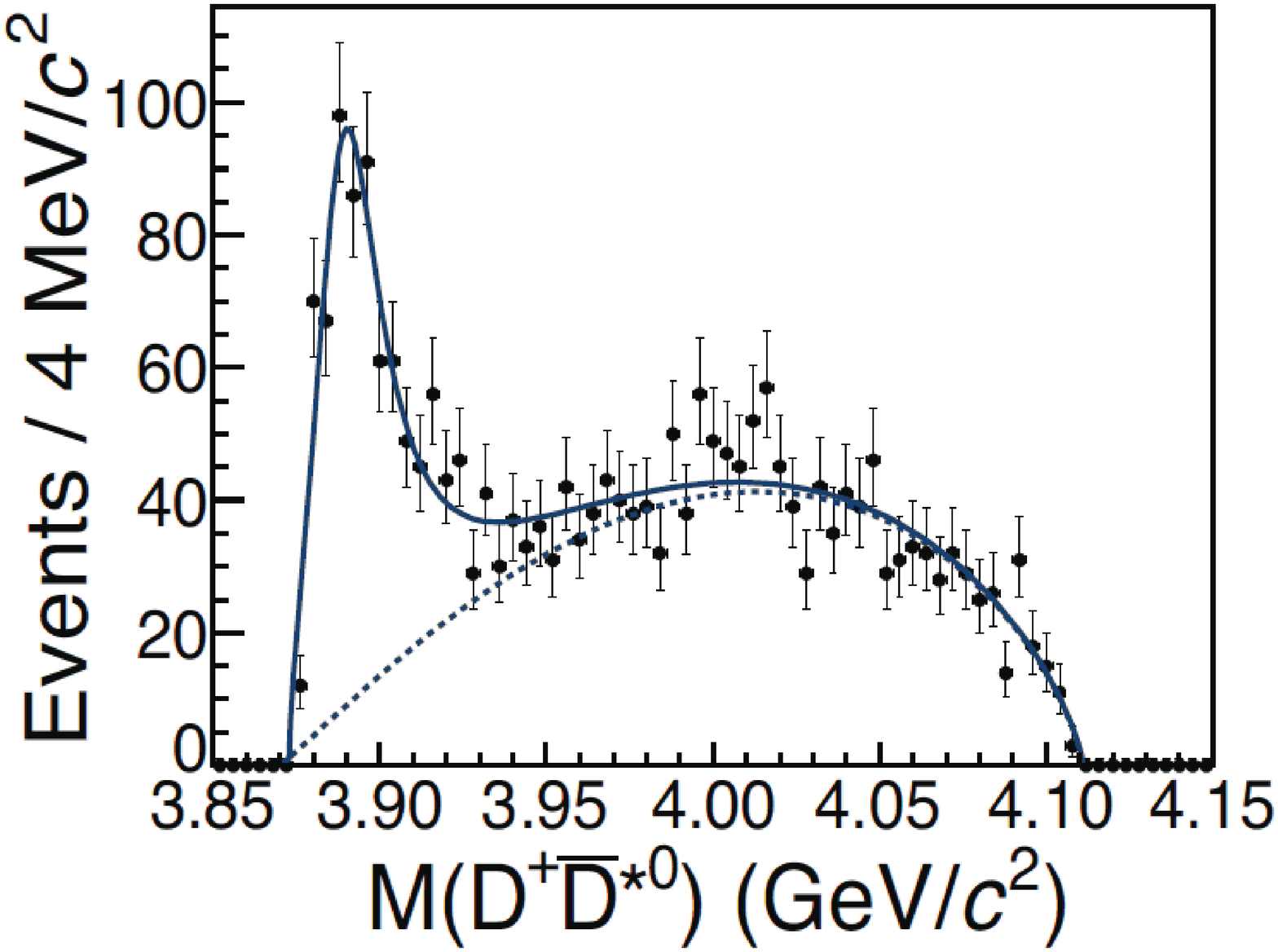}
\end{overpic}
\begin{overpic}[width=0.24\linewidth, height=2.55 cm]{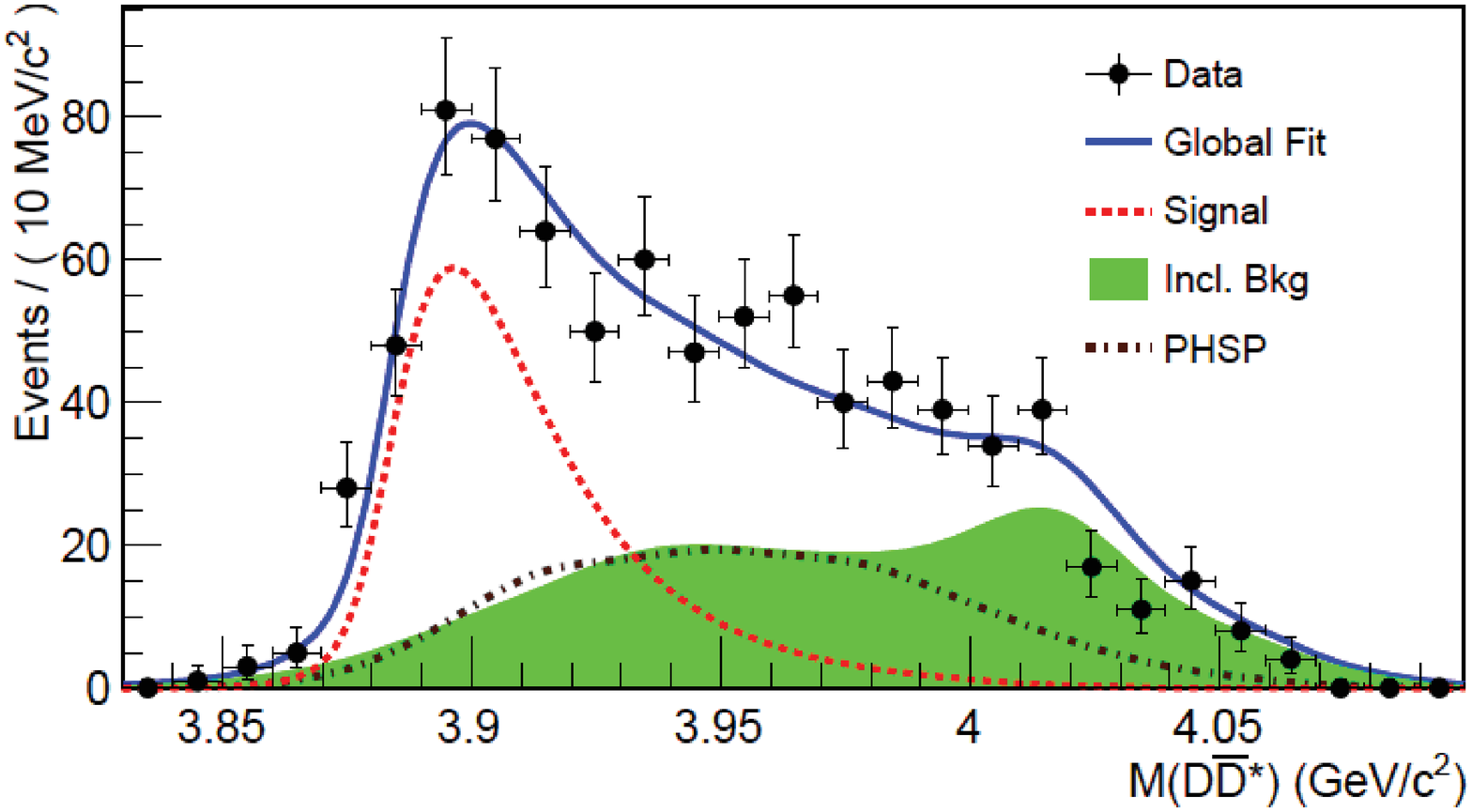}
\end{overpic}
\begin{overpic}[width=0.24\linewidth, height=2.55 cm]{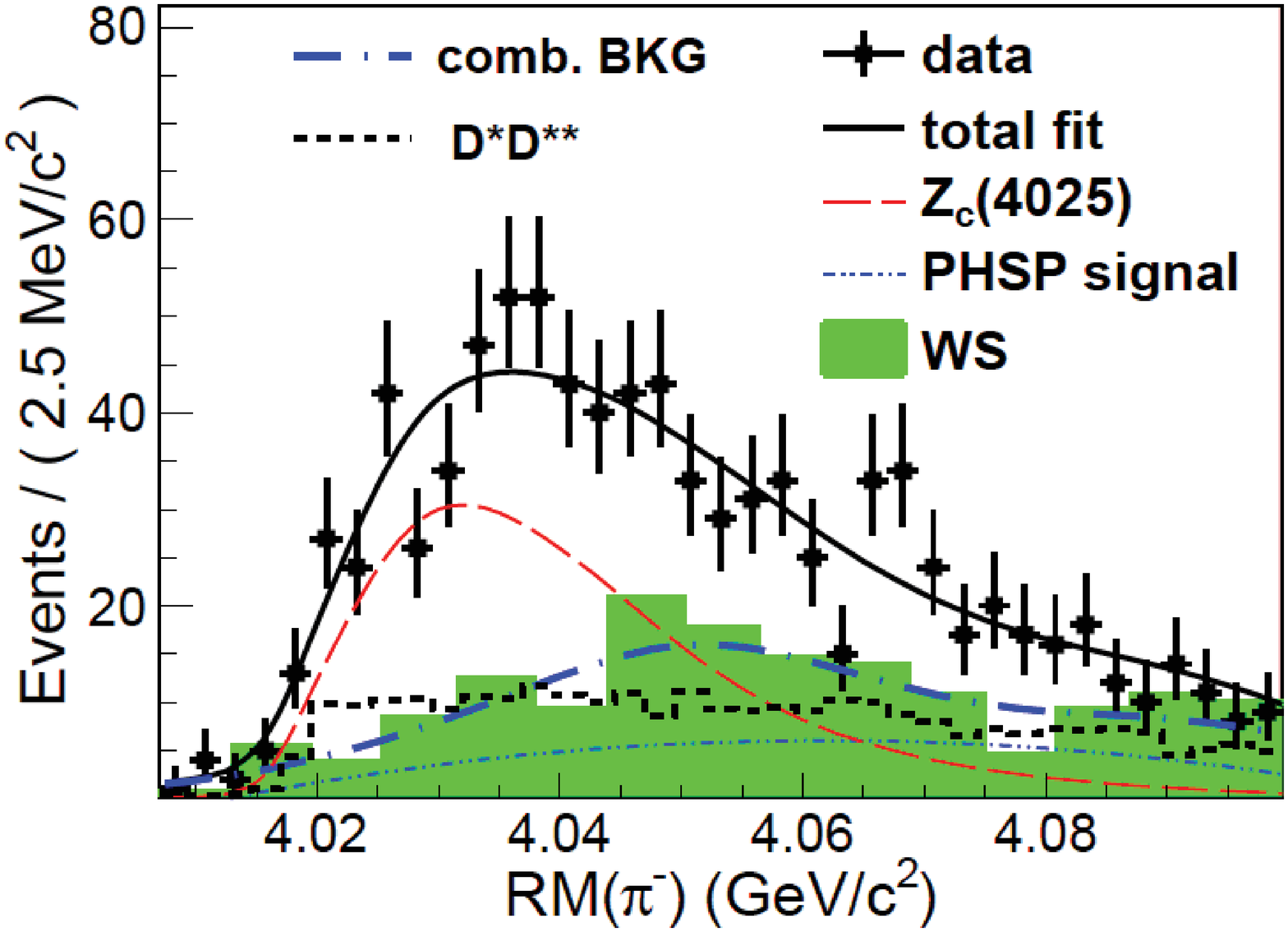}
\end{overpic}
\begin{overpic}[width=0.24\linewidth, height=2.55 cm]{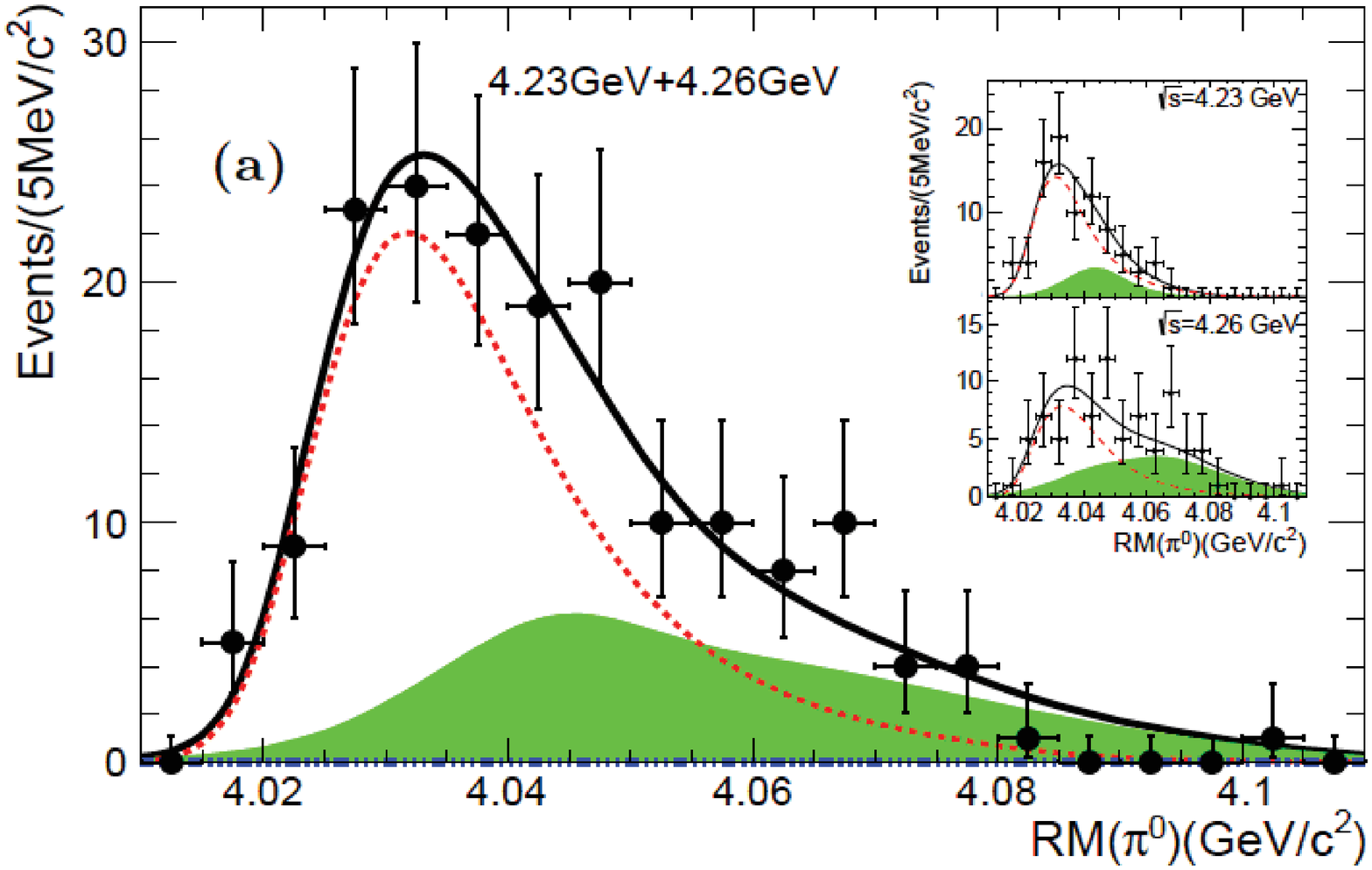}
\end{overpic}
\caption{$Z_c$ states at BESIII}
\label{bes3zc}
\end{figure}

BESIII has observed four-quark state candidates $Z_c(3900)^{\pm}$ and $Z_c(4020)^{\pm}$, and their neutral partners $Z_c(3900)^0$ and $Z_c(4020)^0$ in $e^+e^-\to\pi^{+,-,0} Z_c^{-,+,0}$, $Z_c\to \pi^{0,+,-} J/\psi, \pi^{0,+,-} h_c$ processes. In processes $e^+e^-\to \pi D(D*)\bar{D}*$, BESIII observed $Z_c(3885)$ and $Z_c(4025)$, with masses close to $Z_c(3900)$ and $Z_c(4020)$. These could be considered as the same states~\cite{Ablikim:2013wzq, Ablikim:2013mio, Ablikim:2015tbp,Ablikim:2013xfr, Ablikim:2015gda, Ablikim:2013emm, Ablikim:2015vvn}, as strongly suggested by Figure~\ref{bes3zc}. For details on the studies of the $Z_c$ states at BESIII, see the talk given by Dr. Zhentian Sun at CIPANP2015. 

\section{\boldmath Summary}

Lots of progress has been made in the study of heavy exotic hadronic states at BESIII, including observations of $Z_c$ states, observations of $e^+e^-\to\gamma X(3872)$ and $\pi^+\pi^-X(3823)$ and measurements of line shapes for many hidden-charm final states. BESIII is expected to take data until 2022, so more data will be available to improve our understanding of the nature of the $XYZ$ states.

\end{document}

%% file: econfmacros.tex
\def\beq{\begin{equation}}
\def\eeq#1{\label{#1}\end{equation}}
\def\eeqn{\end{equation}}

\def\beqa{\begin{eqnarray}}
\def\eeqa#1{\label{#1}\end{eqnarray}}
\def\eeqan{\end{eqnarray}}

\let\bar=\overbar

\def\Dslash{\not{\hbox{\kern-4pt $D$}}}
\def\dslash{\not{\hbox{\kern-2pt $\del$}}}

\def\msb{{\bar{\ssstyle M \kern -1pt S}}}

